\font\tenfrakturb=eufb10
\font\tenfraktur=eufm10
\font\tenmsbm=msbm10
\font\sevenfrakturb=eufb7
\font\sevenfraktur=eufm7
\font\sevenmsbm=msbm7
\font\fivefrakturb=eufb5
\font\fivefraktur=eufm5
\font\fivemsbm=msbm5
\newfam\bgothicfam
\newfam\gothicfam
\newfam\msbmfam
\textfont\bgothicfam = \tenfrakturb \scriptfont\bgothicfam=\sevenfrakturb
\scriptscriptfont\bgothicfam=\fivefrakturb

\textfont\gothicfam = \tenfraktur \scriptfont\gothicfam=\sevenfraktur
\scriptscriptfont\gothicfam=\fivefraktur

\textfont\msbmfam = \tenmsbm \scriptfont\msbmfam=\sevenmsbm
\scriptscriptfont\msbmfam=\fivemsbm

\def\frak{\tenfraktur\fam\gothicfam}
\def\Bbb{\tenmsbm\fam\msbmfam}

\def\goth{\tenfraktur\fam\gothicfam}
\catcode`@=11
\def\renewcounter#1{\@definecounter{#1}\@ifnextchar[{\@newctr{#1}}{}}
\documentstyle[12pt]{article}
\def\bh{${\Bbb R}^2\times S^2\>$}

\def\brc{\uppercase\expandafter{\romannumeral1\,\,}}
\begin{document}
\begin{flushright}
\bf gr-qc/9611021
\end{flushright}
\begin{flushright}
Published in {\it Int. J. Mod. Phys.} {\bf A12}, No. 19 (1997) 3347--3363
\end{flushright}
\vspace*{0.37truein}
\setcounter{page}{1}
\centerline{\bf PARAMETRIZATION OF U(N)-MONOPOLES ON BLACK HOLES}
\vspace*{0.035truein}
\centerline{\bf BY THE MODULI SPACE OF HOLOMORPHIC VECTOR BUNDLES}
\vspace*{0.035truein}
\centerline{\bf OVER TWO-SPHERE AND BLACK HOLE ENTROPY}
\vspace*{0.37truein}
\centerline{\bf Yu. P. Goncharov}
\vspace*{0.15truein}
\centerline{\it Experimental Physics Department, State
Technical University}
\baselineskip=10pt
\centerline{\it St. Petersburg 195251, Russia}
\vspace*{0.225truein}
\begin{abstract}{We discuss how to describe $U(N)$-monopoles on the
Schwarzschild and Reissner-Nordstr\"om black holes by the parameters of
the moduli space of holomorphic vector bundles over $S^2$. For $N=$ 2,3 we
obtain such a description in an explicit form as well as the expressions
for the corresponding monopole masses. This gives a possibility to adduce
some reasonings in favour of existence of both a {\it fine structure} for
black holes and the statistical ensemble tied with it which might generate
the black hole entropy. Also there arises some analogy with the famous
$K$-theory in topology.
}{}{}
\end{abstract}
\section{Introductory Remarks}
\vspace*{-0.5pt}
\noindent

  The present paper is a natural continuation of our previous work
of Ref.\cite{Gon96}, so
we shall not dwell upon the motivation of studying the topics being
considered here so long as it has been done in Ref.\cite{Gon96}.
It should be here only recalled
that one of the motivations of writing Ref.\cite{Gon96} was the search for
the additional quantum numbers (nonclassical hair) characterizing black
holes that might help in building a statistical ensemble necessary to
generate the black hole entropy.

  For this purpose in Ref.\cite{Gon96} with the help of the
classification of complex vector bundles over $S^2$ and the Grothendieck
splitting theorem a number of
infinite series of $U(N)$-magnetic monopoles at $N\geq1$ were constructed
in an explicit form on the Schwarzschild and Reissner-Nordstr\"om
black holes.
Also the masses of the given monopoles were estimated to
show that they might reside in black holes as quantum objects. Any
applications to the problem of statistical substantiation of the black hole
entropy have not, however, been given in Ref.\cite{Gon96}.

  It should be noted that the considerations of Ref.\cite{Gon96} employed
only the differentiable structure of complex vector $U(N)$-bundles over
the chosen class of black holes. Therefore, the additional quantum
numbers (topological charges) parametrizing $U(N)$-monopoles were
obtained as the Chern numbers of the
bundles under discussion. It is known, however, (see, e. g.,
Ref.\cite{Okon80}) that the given bundles admit holomorphic structures
whose moduli space can be actually obtained from the Grothendieck
splitting theorem\cite{{Okon80},{Grot56}}. Accordingly, it is of
significant interest to get some description of the mentioned
$U(N)$-monopoles in dependence of the parameters of the moduli space in
question, so long as, in this way, we obtain a marked increase of the
additional quantum numbers characterizing black holes.

  The given paper will be devoted to developing all the mentioned above
questions. After general considerations in Sec. 2 for arbirary $N\geq1$,
the generic scheme is concretized in Sec. 3 for $U(2)$-monopoles, while
in Sec. 4 for $U(3)$-monopoles. Sec. 5 contains the estimations of the
corresponding monopole masses. In Sec. 6 we adduce some reasonings in
favour of existence of both a {\it fine structure} for black holes which is
related with $U(N)$-monopoles and a statistical ensemble for generating
black hole entropy. Also we discuss an analogy arising in the 4$D$ black
hole physics and reminding us the famous $K$-theory in topology. Finally,
Sec. 7 contains concluding remarks, in particular, concerning the
higher-dimensional black holes, while Appendix for inquiry adduces some
facts about functions employed in Sec. 2.

  We write down the black hole metrics under discussion (using
the ordinary set of local coordinates $t,r,\vartheta,\varphi$,
covering all the spacetime background manifold
of the 4$D$ black hole physics
\bh except for a set of the zero measure) in the form
$$ds^2=g_{\mu\nu}dx^\mu\otimes dx^\nu\equiv
C\,dt^2-C^{-1}\,dr^2-r^2(d\vartheta^2+
\sin^2\vartheta\,d\varphi^2) \eqno(1.1)$$
with $C=1-2M/r$ for the Schwarzschild case and $C=1-2M/r+Q^2/r^2$ for the
Reissner-Nordstr\"om case, where $M$, $Q$ are, respectively, a black hole
mass and an electric charge. Besides in generally $r_+\leq r<\infty$ with
$r_+=M+\sqrt{M^2-Q^2}$.

Under the circumstances we have the spatial
part of the metric (1.1) defined on ${\Bbb{R}}\times S^2$-topology as
$$d\sigma^2=C^{-1}\,dr^2+r^2(d\vartheta^2+\sin^2\vartheta\,d\varphi^2)\equiv
\gamma_{ij}\,dx^i\otimes dx^j \eqno(1.2)$$
with $\sqrt{\gamma}=r^2\,\sin\vartheta\,C^{-1/2}=\sqrt{\det(\gamma_{ij})}$.

  Throughout the paper we employ the system of units with $\hbar=c=G=1$,
unless explicitly stated. Finally, we shall denote $L_2(B)$ the set of
the modulo square integrable complex functions on $B$ for any manifold $B$
furnished with an integration measure.

\section{General Considerations}
\vspace*{-0.5pt}
\noindent

  According to Ref.\cite{Gon96} in order to obtain the infinite families of
$U(N)$-monopoles for $N\geq1$, we should
use the Grothendieck splitting theorem \cite{{Okon80},{Grot56}} which asserts
that any complex vector bundle over $S^2$ ( and, as a consequence, over
\bh) of rank $N\geq1$ [i. e., with the structural group
$U(N)$] is a direct sum of $N$ suitable complex line bundles over $S^2$.
  The standard results of algebraic topology (see, e. g., Ref.\cite{Hus66})
say that
$U(N)$-bundles over $S^2$ are in one-to-one correspondence with elements of
the fundamental group of $U(N)$, $\pi_1[U(N)]$. On the other hand, in virtue
of the Bott periodicity $\pi_1[U(N)]={\Bbb{Z}}$ at $N\geq1$ and,
as a result, there exists the countable number of nontrivial complex vector
bundles of any rank $N>1$ over \bh. The sections of such bundles can be
qualified as topologically inequivalent configurations (TICs) of
$N$-dimensional complex scalar field. The above
classification confronts some $n\in{\Bbb{Z}}$ with each $U(N)$-bundle over
\bh-topology. In what follows we shall call it the Chern number of the
corresponding bundle. TIC with $n=0$ can be called {\it untwisted} one
while the rest of the TICs with $n\not=0$ should be referred to as
{\it twisted}.

  So far we tacitly implied that the $U(N)$-bundles were supposed to be
differentiable. Really, they admit holomorphic structures and since each
differentiable complex line bundle over $S^2$ admits only one holomorphic
structure (i. e., the holomorphic and differentiable classifications of
complex line bundles over $S^2$ coincide \cite{Okon80}) then the Grothendieck
splitting theorem in fact gives a description of the moduli space of
$N$-dimensional holomorphic complex vector bundles over $S^2$. Namely, each
$N$-dimensional holomorphic complex vector bundle over $S^2$ is defined by
the only $N$-plet of integers $(r_1,r_2,\ldots,r_N)\in{\Bbb{Z}}^N$,
$r_1\geq r_2\geq\ldots\geq r_N$. Two of such $N$-plets $(r_i)$ and
$(r_i^\prime)$ define the same
differentiable $N$-dimensional bundle if and only if $\sum\limits_i\,r_i=
\sum\limits_i\,r_i^\prime$.
 In Ref.\cite{Gon96} we neglected the above holomorphic structures and, in
consequence, we chose the
$N$-plet characterizing a $N$-dimensional complex bundle with the Chern
number $n\in{\Bbb{Z}}$ in the form $(n,0,\ldots,0)$. Let us now take into
account the existence of the given holomorphic structures, so we should
consider all the above $N$-plets ($r_i$), each $N$-plet representing the
point of the moduli space of $N$-dimensional complex vector bundles
over $S^2$.

  As was shown in Ref.\cite{Gon94}, each complex line bundle (with the
Chern number $n$) over \bh with
the metric (1.1) has a complete set of sections in
$L_2($\bh$)$, so using the fact that all the $U(N)$-bundles over \bh can
be trivialized over the bundle chart of local coordinates ($t,r,\vartheta,
\varphi$) covering almost the whole manifold \bh, the mentioned set can be
written on the given chart in the form
$$f_{\omega lm}={1\over r}e^{i\omega t}R_{\omega l}(r)e^{im\varphi}
P^l_{mn}(\vartheta)\,,
l=|n|,|n|+1,\ldots,|m|\leq l\,, \eqno(2.1)$$
where the explicit form and some properties of the functions
$P^l_{mn}(\vartheta)$
can be found in Appendix (see also Ref.\cite{Gon94}), but we shall not need
them further. It should be
noted that in physical literature devoted to the Dirac monopoles (see, e. g.,
Refs.\cite{Tam31}), the combinations $e^{im\varphi}P^l_{mn}(\vartheta)
=Y_{nlm}
(\vartheta,\varphi)$ are called the {\it monopole (spherical) harmonics}
[that coincide with the ordinary ones at $n=0$, i. e.,
$Y_{0lm}(\vartheta,\varphi)=Y_{lm}(\vartheta,\varphi)]$. As to the functions
$R_{\omega l}(r)$ then they obey the equation
$${d\over dr}\left(Cr^2{d\over dr}{R_{\omega l}\over r}\right)+
(k^2+r^2\omega^2\,C^{-1}){R_{\omega l}\over r}= 0\eqno(2.2) $$
with $k^2=-l(l+1)$, $l=|n|, |n|+1,\ldots$

  Now, in accordance with the Grothendieck splitting theorem, any section of
$N$-dimensional complex bundle $\xi_n$ over \bh with the Chern number
$n\in{\Bbb{Z}}$ can be represented by a $N$-plet ($\phi_1,\ldots,\phi_N$) of
complex scalar fields $\phi_i$, where each $\phi_i$ is a section of a complex
line bundle over \bh. According to the above, we can consider $\phi_i$ the
section of complex line bundle with the Chern number $r_i\in{\Bbb{Z}}$,
where the numbers $r_i$ are subject to the conditions
$$r_1\geq r_2\geq\ldots\geq r_N\>,$$
$$ r_1+r_2+\cdots+r_N=n\>.\eqno(2.3)$$
 As a consequence, we can require the $N$-plets
$$({1\over r}e^{i\omega_1t}R_{\omega_1l_1}(r)Y_{r_1l_1m_1},
   {1\over r}e^{i\omega_2t}
R_{\omega_2l_2}(r)Y_{r_2l_2m_2},\ldots,
   {1\over r}e^{i\omega_Nt}R_{\omega_Nl_N}(r)
Y_{r_Nl_Nm_N})$$ to form the basis in $[L_2($\bh$)]^N$ for the sections of
$\xi_n$, $l_i=|r_i|,|r_i|+1,\ldots$, $|m_i|\leq l_i$,
and this will define the wave equation for a section $\phi=
(\phi_1,\ldots,\phi_N)$ of $\xi_n$ with respect to the metric (1.1)

$$\Biggl[I_N\Box-{1\over r^2\sin^2\vartheta}\times$$

$$\pmatrix{
2ir_1\cos\vartheta\partial_\varphi -r_1^2&0&\ldots&0\cr
0&2ir_2\cos\vartheta\partial_\varphi -r_2^2&\ldots&0\cr
\ldots&\ldots&\ldots&\ldots\cr
0&0&\ldots&2ir_N\cos\vartheta\partial_\varphi-r_N^2\cr}\Biggr]
\pmatrix{\phi_1\cr\phi_2\cr\vdots\cr\phi_N\cr}=0\>,
\eqno(2.4)$$

where $I_N$ is the unit matrix $N\times N$, $\Box =(\delta)^{-1/2}
\partial_\mu(g^{\mu\nu}(\delta)^{1/2}\partial_\nu)$ --- the conventional
wave operator conforming to metric (1.1) with the module of its determinant
$\delta=r^4\sin^2\vartheta$.

  The Eq. (2.4) will, in turn, correspond to the lagrangian
$${\cal L}=\delta^{1/2}g^{\mu\nu}\overline{{\cal D_\mu}\phi}
{\cal D}_\nu\phi\>,\eqno(2.5)$$
 with
$\phi=(\phi_i)$ and a covariant derivative ${\cal D}_\mu=
\partial_\mu-igA^a_\mu\,T_a$ on sections of the bundle $\xi_n$, while the
line
in (2.5) signifies hermitean conjugation and the matrices $T_a$ will form
a basis of the Lie algebra of $U(N)$ in $N$-dimensional space, $a=1,\ldots,
N^2$, $g$ is a gauge coupling constant, i. e., we come to a theory describing
the interaction of a $N$-dimensional twisted complex scalar field with the
gravitational field described by metric (1.1). The coefficients $A^a_\mu$ will
represent a connection in the given bundle $\xi_n$ and will describe some
nonabelian $U(N)$-monopole.

  As can be seen, the Eq.(2.4) has the form ${\cal D^\mu}{\cal D}_
\mu\phi=0$, where ${\cal D}^\mu$ is a formal adjoint to
${\cal D}_\mu$ with regards to the scalar product induced by metric (1.1)
in $[L_2($\bh$)]^N$. That is, the operator ${\cal D}^\mu$ acts on the
differential forms $a_\mu dx^\mu$ with coefficients in the bundle $\xi_n$ in
accordance with the rule

$${\cal D}^\mu(a_\mu)=-{1\over\sqrt{\delta}}\partial_\mu(g^{\mu\nu}
\sqrt{\delta}a_\nu)
+ig\overline{A_\mu}g^{\mu\nu}a_\nu\>\eqno(2.6)$$
with $A_\mu=A^a_\mu T_a$.

 As a result, the equation ${\cal D}^\mu{\cal D}_\mu\phi=0$ takes the form
$$I_N\Box\phi-{1\over\sqrt{\delta}}ig\partial_\mu(g^{\mu\nu}\sqrt{\delta}
A_\nu\phi)-(ig\overline{A_\mu}g^{\mu\nu}\partial_\nu+
g^2g^{\mu\nu}\overline{A_\mu}A_\nu)\phi=0\>.\eqno(2.7)$$

 Comparing (2.4) with (2.7) gives a row of the (gauge) conditions:

\subsection{}
\noindent
$A_t=A_r=0$

\subsection{}
\noindent
$A^a_\vartheta=-\overline{A^a_\vartheta}$ is pure imaginary,
since we, as is accepted in physics, consider the matrices $T_a$ hermitean.

\subsection{}
\noindent
$A^a_\varphi=\overline{A^a_\varphi}\,$ and
$$gA^a_\varphi T_a=-\pmatrix{r_1\cos\vartheta&0&\ldots&0\cr
0&r_2\cos\vartheta&\ldots&0\cr
0&0&\ldots&0\cr
0&0&\ldots&r_N\cos\vartheta\cr}\>,\eqno(2.8)$$
so, accordingly, $A^a_\varphi=A^a_\varphi(\vartheta)$.

\subsection{}
\noindent
$${i\over g}\left[\cot\vartheta\,A^a_\vartheta T_a+(\partial_\vartheta
A^a_\vartheta)T_a\right]+\overline{(A^a_\vartheta T_a)}
(A^b_\vartheta T_b)=$$

$$=\left({1\over g}\right)^2
\pmatrix{ r_1^2 &0     &\ldots &0     \cr
          0     &r_2^2 &\ldots &0     \cr
          0     &0     &\ldots &0     \cr
          0     &0     &\ldots &r_N^2 \cr}
                                           \>, \eqno(2.9)$$

where $A^a_\vartheta=A^a_\vartheta(\vartheta)$ depends only on $\vartheta$
and, as a consequence, the connection matrix $A$ for $\xi_n$-bundle is
equal to
$A=A^a_\mu T_a dx^\mu=A^a_\vartheta(\vartheta) T_a d\vartheta+
A^a_\varphi(\vartheta)T_a d\varphi$ with the $A^a_{\vartheta,\varphi}$
subject to the above conditions.

  This yields the curvature matrix $F=dA+A\wedge A$ for $\xi_n$-bundle
in the form
$$F= F^a_{\mu\nu}T_a dx^\mu\wedge dx^\nu=(\partial_\vartheta A^a_\varphi)
T_a d\vartheta\wedge d\varphi=$$
$$={1\over g}\sin\vartheta d\vartheta\wedge d\varphi
\pmatrix{r_1 &0  &\ldots&0  \cr
         0   &r_2&\ldots&0  \cr
         0   &0  &\ldots&0  \cr
         0   &0  &\ldots&r_N\cr}\>, \eqno(2.10)$$

because the exterior differential $d=\partial_t dt+\partial_r dr+
\partial_\vartheta d\vartheta+\partial_\varphi d\varphi$ in coordinates
$t,r,\vartheta,\varphi$.

 From here it follows that the first Chern class $c_1(\xi_n)$ of the bundle
$\xi_n$ can be chosen in the form

$$ c_1(\xi_n)={g\over4\pi}{\rm Tr}(F)={n\over4\pi}\sin\vartheta
d\vartheta\wedge d\varphi\>,\eqno(2.11)$$

where we employed (2.3) and (2.10), so that if integrating $c_1(\xi_n)$
over topological two sphere $S^2$ (which can be described by the relations
$0\leq\vartheta<\pi$, $0\leq\varphi<2\pi$ in the manifold in question)
we have

$$ \int\limits_{S^2}\,c_1(\xi_n)=n \>, \eqno(2.12)$$
which is equivalent to the conventional Dirac charge quantization condition
$qg=4\pi n$ with (nonabelian) magnetic charge

$$q=\int\limits_{S^2}\,{\rm Tr}(F)\>.\eqno(2.13)$$

  Introducing the Hodge star operator $\ast$ conforming metric (1.1)
on 2-forms $F= F^a_{\mu\nu}T_adx^\mu\wedge dx^\nu$ with the values
in the Lie algebra of $U(N)$ by the relation (see, e. g.,
Refs.\cite{Bes81})

$$(F^a_{\mu\nu} dx^\mu\wedge dx^\nu)\wedge
(\ast F^a_{\alpha\beta} dx^\alpha\wedge
dx^\beta)=g^{\mu\alpha}g^{\nu\beta}F^a_{\mu\nu}F^a_{\alpha\beta}
\sqrt{\delta}\,dx^0\wedge\cdots\wedge dx^3 \>,\eqno(2.14)$$
written in local coordinates $x^\mu$ (there is no summation over $a$
in (2.14)), in coordinates $t,r,\vartheta,\varphi$ we have for $F$ of (2.10)

$$\ast F=\ast F^a_{\mu\nu}T_a dx^\mu\wedge dx^\nu=
{1\over g}r^{-2}dt\wedge dr
\pmatrix{r_1&0&\ldots&0\cr
         0&r_2&\ldots&0\cr
         0&  0&\ldots&0\cr
         0&  0&\ldots&r_N\cr}\>.\eqno(2.15)$$

We can now consider the Yang-Mills (Maxwell at $N=1$) equations

$$dF=F\wedge A - A\wedge F \>, \eqno(2.16)$$
$$d\ast F= \ast F\wedge A - A\wedge\ast F \>.\eqno(2.17)$$

It is clear that (2.16) is identically satisfied by the above $A, F$. As
for the Eq. (2.17) that it reduces to the condition

$$ A^a_\vartheta(\vartheta)T_a
\pmatrix{r_1&0&\ldots&0\cr
         0&r_2&\ldots&0\cr
         0&  0&\ldots&0\cr
         0&  0&\ldots&r_N\cr}=
\pmatrix{r_1&0&\ldots&0\cr
           0&r_2&\ldots&0\cr
           0&  0&\ldots&0\cr
           0&  0&\ldots&r_N\cr}A^a_\vartheta(\vartheta)T_a\eqno(2.18)$$

which can be fulfilled (at $N=1$ always), for example, if the matrix
$A^a_\vartheta T_a$ is
diagonal, so there exist nontrivial solutions of (2.18).

  To evaluate the monopole masses we should use the $T_{00}$-component
of the energy-momentum tensor

$$T_{\mu\nu}={1\over4\pi}(-F^a_{\mu\alpha}F^a_{\nu\beta}g^{\alpha\beta}
+\frac{1}{4}
F^a_{\beta\gamma}F^a_{\alpha\delta}\,g^{\alpha\beta}g^{\gamma\delta}
g_{\mu\nu})\eqno(2.19)$$
 which does obviously not depend on $A^a_\vartheta(\vartheta)$, hence
the main thing
is the solutions of (2.8). Having obtained them we can find the monopole
masses according to

$$m_{\rm{mon}}(n)=\int\limits_{{\Bbb R}\times S^2}\,
T_{00}\sqrt{\gamma}d^3x
=\int\limits_{{\Bbb R}\times S^2}
T_{00}\,r^2\sin\vartheta\,C^{-1/2}d^3x\>.\eqno(2.20)$$

with $C$ of (1.1) and $\gamma$ of (1.2) while $T_{00}$-component is
evaluated on
the above solutions. Therefore, let us concretize the above
construction for the cases $N=2,3$ since the $N=1$ case does not differ
from the one of Ref.\cite{Gon96}. But before one should do a remark.

  Generally speaking for arbitrary spacetimes the formula (2.20) is
badly defined since it is impossible in general case to separate the full
energy of a gravitating system into the energy of matter (including the
electromagnetic field) and the energy of gravitational field itself. However,
as was discussed many years ago \cite{Mis73}, there exists a class of
spacetimes where
such a separation is possible. These are the so-called {\it asymptotically
flat spacetimes} (AFS). Moreover, for AFS one can introduce the so-called
the {\it pseudotensor} of Landau-Lifshits $t_{\mu\nu}^{L-L}$ of gravitational
field \cite{{Mis73},{LL67}}, so that the full effective energy-momentum
tensor of the gravitating
system will be $T_{\mu\nu}^{eff}=T_{\mu\nu}+t_{\mu\nu}^{L-L}$ with the
energy-momentum tensor of matter (including electromagnetic field)
$T_{\mu\nu}$.

  If $V$ is a spatial part of the AFS then the quantity
$E=\int_{V}T_{00}^{eff}\sqrt{\gamma}d^3x$ is interpreted
as the full energy of the gravitating system and, clearly, $E$ can be
considered as the sum of the energy of matter and the energy of gravitational
field itself. As is well known \cite{Mis73}, the black hole spacetimes are just
of the AFS and, for example, in Ref.\cite{LL67} the contribution
$\int_{V}t_{00}^{L-L}\sqrt{\gamma}d^3x$ was evaluated for the Schwarzschild
spacetime and proved to be equal to $M$, black hole mass. Therefore, the rest
of the full energy in AFS equal to $\int_{V}T_{00}\sqrt{\gamma}d^3x$ can be
interpreted as the energy (or, in units used by us, as mass) of matter
(including electromagnetic field) and, as a result, the formula (2.20) makes
sense in black hole spacetimes with the interpretation used further in
our paper.

\section{U(2)-monopoles}
\vspace*{-0.5pt}
\noindent

  In this case we can take $T_1=I_2$, $T_a=\sigma_{a-1}$ at $a=2,3,4$, where
$\sigma_{a-1}$ are the ordinary Pauli matrices
$$\sigma_1=\pmatrix{0&1\cr 1&0\cr}\,,\sigma_2=\pmatrix{0&-i\cr i&0\cr}\,,
\sigma_3=\pmatrix{1&0\cr 0&-1\cr}\,\,. \eqno(3.1)$$
  Then the Eq. (2.8) will be consistent with such a choice, if we put the
connection matrix $A=A^a_\mu\,T_a\,dx^\mu$ to be equal to
$$A= -{1\over2g}\cos\vartheta\,d\varphi[(r_1+r_2)I_2+
(r_1-r_2)\sigma_3]+A^a_\vartheta\,T_a
d\vartheta\>, \eqno(3.2)$$
where $A^a_\vartheta(\vartheta)$ possesses the properties described in
Sec. 2.

  This yields the curvature matrix $F=dA+A\wedge A$ in the form
$$F=F^a_{\mu\nu}\,T_a\,dx^\mu\wedge dx^\nu=
{1\over2g}\sin\vartheta d\vartheta\wedge d\varphi[(r_1+r_2)I_2+
(r_1-r_2)\sigma_3]=$$
$${1\over2g}\sin\vartheta d\vartheta\wedge d\varphi[nI_2+
(r_1-r_2)\sigma_3]\>, \eqno(3.3)$$

where we used the relation (2.3).

 In accordance with (2.20) we find

$$T_{00}={1\over16\pi}\left(1-{2M\over r}+{Q^2\over r^2}\right){1\over r^4}
\left({1\over2g}\right)^2[n^2+(r_1-r_2)^2]\>.\eqno(3.4)$$
In what follows we denote $A_2(n)=n^2+(r_1-r_2)^2$.

\section{U(3)-monopoles}
\vspace*{-0.5pt}
\noindent

  In the given situation we can take $T_1=I_3$, $T_a=\lambda_{a-1}$
at $a=2,\ldots,9$, where $\lambda_{a-1}$ are the Gell-Mann matrices
$$\lambda_1=\pmatrix{0&1&0\cr 1&0&0\cr 0&0&0\cr}\,,
  \lambda_2=\pmatrix{0&-i&0\cr i&0&0\cr 0&0&0\cr}\,,
  \lambda_3=\pmatrix{1&0&0\cr 0&-1&0\cr 0&0&0\cr}\,,  $$
$$ \lambda_4=\pmatrix{0&0&1\cr 0&0&0\cr 1&0&0 \cr}\,,
   \lambda_5=\pmatrix{0&0&-i\cr 0&0&0\cr i&0&0\cr}\,,
   \lambda_6=\pmatrix{0&0&0\cr 0&0&1\cr 0&1&0\cr}\,, $$
$$\lambda_7=\pmatrix{0&0&0\cr 0&0&-i\cr 0&i&0\cr}\,,
  \lambda_8={1\over\sqrt3}\pmatrix{1&0&0\cr 0&1&0\cr
                   0&0&-2\cr}\,. \eqno(4.1)   $$
  For the Eq. (2.8) to be consistent with such a choice, it should be put
$$A=A^a_\mu\,T_a\,dx^\mu=-{1\over3g}\cos\vartheta\,d\varphi
[(r_1+r_2+r_3)I_3+{3\over2}(r_1-r_2)\lambda_3+$$
$${\sqrt3\over2}(2r_3-r_1-r_2)\lambda_8]+A^a_\vartheta
\,T_a\,d\vartheta \>.
\eqno(4.2)$$

  This yields
$$F= F^a_{\mu\nu}\,T_a\,dx^\mu\wedge dx^\nu={1\over3g}
\sin\vartheta\,d\vartheta
\wedge d\varphi[(r_1+r_2+r_3)I_3+{3\over2}(r_1-r_2)\lambda_3+$$
$${\sqrt3\over2}(2r_3-r_1-r_2)
\lambda_8]\>. \eqno(4.3)$$

Finally, taking into account (2.3)

$$T_{00}={1\over16\pi}\left(1-{2M\over r}+{Q^2\over r^2}\right){1\over r^4}
\left({1\over3g}\right)^2\times$$
$$[n^2+{9\over4}(r_1-r_2)^2+{3\over4}(2r_3-r_1-r_2)^2]
\>.\eqno(4.4)$$
In what follows we denote $A_3(n)=n^2+{9\over4}(r_1-r_2)^2+{3\over4}
                                       (2r_3-r_1-r_2)^2$.

 It is clear that the case of arbitrary $N$ can be treated analogously but
we shall not dwell upon it.

\section{Monopole Masses}
\vspace*{-0.5pt}
\noindent

 In accordance with (2.20) let us consider miscellaneous cases and for
inquiry we shall also adduce the results for $N=1$ case of Ref.\cite{Gon96}.

\subsection{Schwarzschild black hole}
\noindent
We have $Q=0$, $r_+=2M$.

\subsubsection{$U(1)$-monopoles}
\noindent
$$m_{\rm {mon}}(n)={1\over12M}\left({n\over e}\right)^2 \eqno(5.1)$$
with $g=e=4.8\cdot10^{-10}\,{\rm cm^{3/2}g^{1/2}s^{-1}}$ in usual units.

\subsubsection{$U(2)$-monopoles}
\noindent
$$m_{\rm {mon}}(n,r_1,r_2)=\left({1\over2g}\right)^2{A_2(n)\over4}
\int\limits_{2M}^\infty\,\sqrt{1-{2M\over r}}{dr\over r^2}
=\left({1\over2g}\right)^2{A_2(n)\over12M}\>.\eqno(5.2)$$

\subsubsection{$U(3)$-monopoles}
\noindent
$$m_{\rm {mon}}(n,r_1,r_2,r_3)=\left({1\over3g}\right)^2
               {A_3(n)\over12M}\>.\eqno(5.3)$$

\subsection{Reissner-Nordstr\"om black hole with $Q\ne M$}
\noindent
We have here $r_+=M+\sqrt{M^2-Q^2}$. It is easy to notice that at
$r_+\leq r<\infty$ the quantity $(1-2M/r+Q^2/r^2)\leq1$.

\subsubsection{$U(1)$-monopoles}
\noindent
$$m_{\rm {mon}}(n)={n^2\over4e^2}\int\limits_{r_+}^\infty
\,\sqrt{1-{2M\over r}+{Q^2\over r^2}}\frac{dr}{r^2}\approx
{n^2\over4e^2}\int_{r_+}^\infty\,{dr\over r^2}=
{n^2\over4e^2r_+}\>.\eqno(5.4)$$

\subsubsection{$U(2)$-monopoles}
\noindent
$$m_{\rm {mon}}(n,r_1,r_2)=\left({1\over2g}\right)^2{A_2(n)\over4}
\int\limits_{r_+}^\infty\,\sqrt{1-{2M\over r}+{Q^2\over r^2}}{dr\over r^2}
\approx\left({1\over2g}\right)^2{A_2(n)\over4}\int\limits_{r_+}^\infty
\,{dr\over r^2}$$
$$=\left({1\over2g}\right)^2{A_2(n)\over4r_+}\>.
\eqno(5.5)$$

\subsubsection{$U(3)$-monopoles}
\noindent
$$m_{\rm {mon}}(n,r_1,r_2,r_3)
\approx\left({1\over3g}\right)^2{A_3(n)\over4r_+}
\>.\eqno(5.6)$$

\subsection{Extremal Reissner-Nordstr\"om black hole}
\noindent
We have $Q=M$ and $r_+=M$. Under these conditions we can make replacement
$1/r=x$, $x_+=1/r_+=1/M$ and take into account that
$\sqrt{Q^2x^2-2Mx+1}=M|1/M-x|$ which gives
$$\int\limits_0^{x_+}\sqrt{Q^2x^2-2Mx+1}\,dx={1\over8M}\>.\eqno(5.7)$$
This yields

\subsubsection{$U(1)$-monopoles}
\noindent
$$m_{\rm {mon}}(n)=\left({n\over e}\right)^2{1\over8M}\>. \eqno(5.8)$$

\subsubsection{$U(2)$-monopoles}
\noindent
$$m_{\rm {mon}}(n,r_1,r_2)=\left({1\over2g}\right)^2{A_2(n)\over8M}
\>.\eqno(5.9)$$

\subsubsection{$U(3)$-monopoles}
\noindent
$$m_{\rm {mon}}(n,r_1,r_2,r_3)=\left({1\over3g}\right)^2{A_3(n)\over8M}
\>.\eqno(5.10)$$

 It should be noted that for to obtain the above monopole masses in usual
units it is enough to multiply the values (5.1)--(5.6) and (5.8)--(5.10)
by $\hbar^2c^2/G$. Introducing parametrization $Q=\alpha M$ with
$0\leq\alpha\leq1$, we find $r_+=Mf(\alpha)$ with
$f(\alpha)=1+\sqrt{1-\alpha^2}$, $1\leq f(\alpha)\leq2$.
Under the circumstances, evaluating the corresponding Compton wavelength
$\lambda_{\rm {mon}}(n,r_i)=\hbar/m_{\rm {mon}}(n,r_i)c$, we can see
that at any $n\ne0, N\geq1$, $\lambda_{\rm {mon}}(n,r_i)\ll r_g$, where
$r_g=r_+G/c^2$ is a gravitational radius of black hole, if $g^2/\hbar c\ll1$.
The latter always holds true for $U(1)$-monopoles so long as $e^2/\hbar c
\sim 1/137$. As a consequence, we come
to the conclusion that under certain conditions $U(N)$-monopoles might
reside in black holes as quantum objects.

  So, we can see that the masses of $U(N)$-monopoles really depend on
the parameters of the moduli space of holomorphic vector bundles
over $S^2$. Let us consider some possible issues for the 4$D$ black hole
physics from this fact, having denoted this moduli space (whose
description has been adduced in Sec. 2 ) as ${\goth M}_N$ for some given
$N\geq1$.

\section{Fine Structure of Black Holes, a Statistical Ensemble for
Generating Black Hole Entropy, Quantum $K$-theory}
\vspace*{-0.5pt}
\noindent

   Among the unsolved questions of modern $4D$ black hole physics the
so-called  black hole information problem admittedly ranks high. Referring
for more details, e. g., to Refs.\cite{Gid94} (and references quoted
therein), it should be noted here that one aspect of the problem consists
in that
for an external observer any black hole looks like an object having in general
only a finite number of parameters (classical hair --- mass $M$, charge $Q$,
angular momentum $J$) and it is, therefore, unclear how these parameters
can encode
all the information about quantum particles of matter (which has been
collapsed to the black hole), particles that are being radiated \`a la Hawking.
 As a consequence, it is impossible to distinguish all the black hole (pure)
states, so a black hole should, therefore, be described by a mixed state.
In other words, the system (black hole) has an entropy $S$ while
the latter does not correspond to any statistical ensemble, so long as
there is no infinite number of quantum (discrete) numbers connected with
this system to build an appropriate statistical ensemble.

 As the results of both Ref.\cite{Gon96} and the present paper, however,
show, the natural candidates for additional quantum numbers
(nonclassical hair)
for black holes might be the topological quantum numbers parametrizing
$U(N)$-monopoles on black holes, so these numbers could be identified with
${\goth M}_N$. Really, as has been demonstrated recently in Ref.\cite{Gon95},
black holes can radiate \`a la Hawking for any TICs, for instance, of
complex scalar field with the Chern number $n\in {\Bbb Z}={\goth M}_1$ and
this occurs independently of other field configurations. More exact
analytical and numerical considerations \cite{GF96} show that, for instance,
in the Schwarzschild black hole case, twisted TICs
can give the marked additional contribution of order 17 \% to the total
luminosity (summed up over all the TICs). This tells us that
there exists some {\it fine structure} in black hole physics which is
conditioned by nontrivial topological properties of black holes and the
given fine structure is able to markedly modify the black hole
characteristics, so long as, for example, the words " Hawking radiation for
complex scalar field " should be now understood as the radiation summed up
over all the TICs of complex scalar field on black hole. In a sense,
the black hole fine structure is quite analogous to the one of atomic spectra
in atomic physics where its existence enables us to achieve an essentially
better understanding of the whole structure of atoms.

Let us consider, therefore, more in
detail in which way the above fine structure might help to black holes
to form a statistical ensemble necessary to generate the black hole entropy.

\subsection{Schwarzschild black hole}
\noindent
    As is known (see, e. g., Ref.\cite{Nov86}), the entropy $S$ of black hole
can be introduced from purely
thermodynamical considerations and, for example, $S=4\pi M^2$ in the
Schwarzschild black hole case, so when putting the internal energy of black
hole $U=M$, we obtain the temperature of black hole
$T={\partial U\over\partial S}={1\over8\pi M}$ through the standard
thermodynamical relation. It is obvious that $S$ corresponds to a formal
partition function $Z=e^{-{M\over2T}}$ for the given $M$ and $T$ ( we took the
Boltzmann constant $k_B =1$). The quantity $Z$ is formal because we cannot
point out any infinite statistical ensemble conforming to it, so that one could
obtain $Z$ by the usual Gibbs procedure, i. e., by averaging over
this ensemble.
   The results of Refs.\cite{{Gon95},{GF96}} show that black hole can
radiate \`a la Hawking
for any TIC of complex scalar field with the Chern number $n\in{\goth M}_1$.
Such a radiation is practically
defined by a couple ($g$, $n$) with the black hole metric $g$ of (1.1) and the
Chern number $n$ in the sense that these data are sufficient to describe the
physical quantities (for instance, luminosity $L(n)$) characterizing
the radiation process for TICs with the Chern number $n$
\cite{{Gon95},{GF96}}.
On the other hand,
as is known (see, e. g., Ref.\cite{Bir82}), the
Hawking effect is being obtained when considering the system (black hole +
matter field near it) semiclassically: the black hole is being described
classically while the matter field is being quantized. All mentioned above
suggests that the Hawking process occurs for the given pair
($g$, $n$) when the black hole is in {\it a quantum state} which can be
characterized by the {\it semiclassical} energy
$$E_n={M\over2}+{\cal E}(n)={M\over2}+{1\over12M}\left({n\over e}\right)^2
            \eqno(6.1)$$
with ${\cal E}(n)=m_{\rm {mon}}(n)$ of (5.1), so long as ${\cal E}(n)$
is a natural
energy of the monopole with the Chern number $n$ residing in black hole,
since the additional contribution to the Hawking radiation is conditioned
actually by the same monopole \cite{{Gon95},{GF96}}. We call $E_n$
semiclassical
because the first term of (6.1) in usual units does not depend on $\hbar$
while the second one does (see Sec. 5).

  Under the circumstances there arises an infinite set of quantum states
($g$, $n$) with the energy spectrum (6.1) for black hole. After this, the
Gibbs average takes the form
$$Z=\sum\limits_{n\in{\Bbb{Z}}}e^{-{E_n\over T}}=
e^{-{M\over2T}}\sum\limits_{n\in{\Bbb{Z}}}e^{-{{\cal E}(n)\over T}}=
e^{-4\pi M^2}\sum\limits_{n\in{\Bbb{Z}}}q^{n^2}=
e^{-4\pi M^2}\vartheta_3(0,q)\eqno(6.2)$$
with the Jacobi theta function $\vartheta_3(v,q)$ and
$q=\exp\left(-{2\pi\over3e^2}\right)$. As a result, we obtain an inessential
constant additive correction $S_1=\ln\vartheta_3(0,q)$ independent of $M$ to
the black hole entropy $S=4\pi M^2$ but now the latter is the result of
averaging over an infinite ensemble which should be considered as inherent to
black hole due to its nontrivial topological properties.

  It is clear that one can also consider all the triplets ($g$, $r_1$, $r_2$),
where the pair ($r_1$, $r_2$) parametrizes the moduli space of
$U(2)$-monopoles ${\goth M}_2$
(see Sec. 2 ), so that the Gibbs average should be accomplished over
${\goth M}_2$ which will again lead to some inessential additional correction
to the entropy $S$ due to dependence (5.2). Moreover, this scheme will
obviously hold true for $U(N)$-monopoles at any $N>1$ if the Gibbs average is
accomplished over the moduli space of $U(N)$-monopoles ${\goth M}_N$
(see Sec. 2 ).

\subsection{Reissner-Nordstr\"om black hole with $Q\ne M$}
\noindent
 The formal partition function for the given case is
$$Z=\exp\left(-{M\over T}\left(1-{\sqrt{1-\alpha^2}\over2}\right)\right)
\eqno(6.3)$$
with $T=\sqrt{1-\alpha^2}/[2\pi M(1+\sqrt{1-\alpha^2})]$. Under the
circumstances the energy spectrum for black hole which is tied, for
example, with $U(1)$-monopoles can be chosen according to (5.4) in the
form
$$E_n\approx M\left(1-{\sqrt{1-\alpha^2}\over2}\right)+
m_{\rm{mon}}(n){\sqrt{1-\alpha^2}\over1+\sqrt{1-\alpha^2}}\eqno(6.4)$$
with $m_{\rm{mon}}(n)$ of (5.4), so again we shall get an inessential
constant additive correction independent of $\alpha$ to the entropy
$S=\pi r_+^2$ after accomplishing
the Gibbs average over the moduli space ${\frak M}_1$. Obviously, the same
holds true for any $U(N)$-monopoles provided that we shall accomplish the
Gibbs average over the moduli space ${\frak M}_N$.

\subsection{Extremal Reissner-Nordstr\"om black hole}
\noindent
  In this case the monopole massess are well defined and exist
(see (5.8)--(5.10)), but we face the general difficulty of defining the
entropy $S$ and temperature $T$ for this extremal case. At present there is no
generally accepted consistent definition for the given quantities though
in literature there have been done many attempts of analysing this situation
(see, e. g., Ref.\cite{Gh95}). We shall not, therefore, dwell upon
the given case.

 Finally, as for the general Kerr-Newman
case, at our disposal there are not yet any expressions of monopole masses in
dependence of black hole parameters $M$, $Q$, $J$ for this case, so that the
reasonings of the given section should be specified for the latter case after
evaluating the necessary quantities that without doubt exist
(see Ref.\cite{Gon96}).

     As is clear from all the above, the black hole topology can bring many
possibilities for producing a huge amount of new quantum numbers. So far we
have, however, mainly spoken about bosons. As to the TICs of fermions, it is
known \cite{{Bes81},{Spin89}} that the given topology admits a countable
number of the
so-called Spin$^c$-structures and this might generate a range of new quantum
charges for fermions on black holes. It should be noted that there exists
topological duality between TICs of real scalar fields and usual spinorial
structures (and the corresponding spinor fields) in the sense that both
classes are classified by the same cohomology group
$H^1(B,{\Bbb{Z}}_2)$, the first cohomology group with coefficients in
${\Bbb{Z}}_2$ for the given manifold $B$, and this duality has nontrivial
applications in quantum
field theory, cosmology and $p$-branes (see our review of Ref.\cite{G94}).
On the
other hand, the classifying group for Spin$^c$-structures is
$H^2(B,{\Bbb{Z}})$, the second cohomology group with coefficients in
${\Bbb{Z}}$ for manifold $B$, that is, the same as for TICs of complex
scalar fields on the manifold $B$ (see, e. g., Refs.\cite{{Gon94},{Bes81}}).
For manifolds underlying the 4$D$ black hole physics this group is equal to
${\Bbb{Z}}$\cite{{Gon96},{Gon94}}, so we obviously in the 4$D$ black hole
physics deal with complex analog of the above duality.

It seems to us, all the mentioned possibilities
should be investigated from physical point of view, in particular, the
influence
of topological quantum numbers on quantum effects for fields near black holes.
It is quite plausible that such a study will allow us to come to the conclusion
that in black hole physics we deal with some (quantum) analog of the famous
$K$-theory in algebraic topology. One can recall that $K$-theory takes into
account {\it all} vector bundles over one or another manifold to build an
appropriate topological
invariant ($K$-ring) for manifolds (see, e. g., Refs.\cite{{Hus66},{Kar78}}).
At this, however,
there exist a number of bundles which are more important for constructing
$K$-ring ( and they, in essence, define it) than other ones over the given
manifold. Perhaps, in black hole physics also topological quantum numbers
which are tied with nontrivial vector bundles over the black hole topology
can be split into more important and less important ones. The former could,
for example, correspond to the observed physical fields, the latter could be
directly unobserved but might help to build a statistical ensemble necessary
to generate black hole entropy. But both classes, at any rate, should
probably be certain relics from quantum gravity processes within black holes.

\section{Concluding Remarks}
\vspace*{-0.5pt}
\noindent

 The results of both the present paper and
Refs.\cite{{Gon96},{Gon94},{Gon95},{GF96}} show that
the 4$D$ black hole physics can have a rich fine structure connected
with the topology \bh underlying the 4$D$ black hole spacetime manifolds.
It seems to be quite probable that this fine structure could manifest
itself in solving the whole number of problems within black hole physics
so that one should seemingly thoroughly study the arising possibilities.

In view of all the above, one can a little touch upon the
miscellaneous attempts of modelling the 4$D$ black hole physics
by considering the $D=2$ and $D>4$-cases that are at present popular enough
in literature.

 Generally speaking, from our point of view, the 2$D$ black hole physics is
plausible not very good laboratory for modelling the 4$D$ case. Indeed, as
a rule, the solutions describing a black hole within various 2$D$ theories
(see, e. g., Refs.\cite{Wit91}) are defined on trivial ${\Bbb{R}}^2$ or
semitrivial ${\Bbb{R}}\times S^1$ background and, accordingly, lose essential
topologial features of the 4$D$ case.

   In contrast to $2D$ case, the one of higher dimensions is seemingly
far more
interesting. Really, in a number of the gravitation theories one has been
demonstrated (see, e. g., Refs.\cite{Bou85}) that there exist the solutions
describing black
holes in any dimension $D>2$. These black hole solutions are naturally defined
on the manifolds with topologies ${\Bbb{R}}^2\times S^{D-2}$, i. e., they are
some reasonable extensions of the $4D$ black hole solutions, possess
thermodynamic properties and, as a result, an entropy. The black hole
information
problem can, therefore be posed for these solutions as well. But it is clear
that the given solutions can also carry nonclassical hair in the sense
described
in the present paper. Indeed, it is obvious that
the ${\Bbb{R}}^2\times S^{D-2}$-topologies admit a huge number of nontrivial
complex and real vector bundles whose classification is evidently reduced to
that for the $n$-sphere $S^n$, $n=D-2$. As to the latter, standard results of
algebraic topology (see, e. g., Ref.\cite{Hus66}) say that the $G$-vector
bundles over $S^n$
for any Lie group $G$ are classified by $\pi_{n-1}(G)$, the $(n-1)$-th
homotopic
group of $G$. On the other hand, by virtue of the famous Bott
periodicity \cite{Bot59}, we
have $\pi_{2k-1}(U(m))={\Bbb{Z}}$, $m>1$, $k=1,2,\ldots$ and, as a consequence,
over each even-dimensional sphere there exist a huge number of nontrivial
complex vector bundles of any rank $m>1$. As for the odd-dimensional spheres
then, according to the Bott periodicity, $\pi_{2k}(U(m))=0$, i. e., no
nontrivial
complex vector bundles exist and, besides, in accordance with the Bott
periodicity for the orthogonal group $O(m)$, $\pi_{2k}(O(m))=0$ as well,
i. e.,
there are not nontrivial real vector bundles either. As a result, we have the
hair of type described in the present paper for black holes of
Refs.\cite{Bou85}
with topologies ${\Bbb{R}}^2\times S^{2k}$, $k=1,2,\ldots$

  In conclusion, it should be noted that some part of the above has been told
by the author within the framework of scientific meeting \cite{Con96}.
\vskip2.0cm
\centerline{\bf Acknowledgement}
\vskip0.5cm
\noindent
    The work was supported in part by the Russian Foundation for
Basic Research (the grant no. 95-02-03568-a).
\vskip0.5cm
\noindent
\centerline{\bf Appendix}
\vskip0.5cm
\noindent
 We give here for inquiry some information about functions
$P^l_{mn}(\cos\vartheta)$ mentioned in Sec. 2.
 The explicit form of the functions $P^l_{mn}(\cos\vartheta)$
can be chosen by miscellaneous ways, for instance, as follows (see, e. g.,
Ref. \cite{Vil91})
$$P^l_{mn}(\cos\vartheta)=(-1)^{-m}i^{-n}
\sqrt{{(l-m)!(l-n)!\over(l+m)!(l+n)!}}
\cot^{m+n}{\vartheta\over2}\,\times$$
$$\times\sum\limits_{k={\rm{max}}(m,n)}^l
{(l+k)!i^{2k}\over(l-k)!(k-m)!(k-n)!}\sin^{2k}{\vartheta\over2}\>,
\eqno({\rm A}.1)$$
which holds true for integer and half-integer $l,n$, $|n|\leq l$.
There is an orthogonality relation at $n$ fixed
$$\int\limits_0^\pi\,\overline{P^l_{mn}}(\cos\vartheta)
P^{l^\prime}_{m^\prime n}(\cos\vartheta)
\sin\vartheta d\vartheta={2\over2l+1}\delta_{ll^\prime}
\delta_{mm^\prime}\>.\eqno({\rm A}.2)$$

$P^l_{mn}(\cos\vartheta)$ obeys the differential equation
$${d\over dx}\left[(1-x^2){d\over dx}\right]P^l_{mn}(x)-
{m^2+n^2-2mnx\over1-x^2}P^l_{mn}(x)=
-l(l+1)P^l_{mn}(x)\eqno({\rm A}.3)$$
with $x=\cos\vartheta$.

As was mentioned in Sec. 2, in physical literature devoted to the Dirac
monopoles (see, e. g.,
Refs. \cite{Tam31}), the combinations $e^{im\varphi}P^l_{mn}(\cos\vartheta)=
Y_{nlm}
(\vartheta,\varphi)$ are called the {\it monopole (spherical) harmonics}
that coincide with the ordinary ones at $n=0$, i. e.,
$Y_{0lm}(\vartheta,\varphi)=Y_{lm}(\vartheta,\varphi)$, that is, the
relations (A.1)--(A.3) pass on to the standard relations for ordinary spherical
harmonics \cite{Vil91}. It should be
noted, however, that in mathematical literature the monopole harmonics
have been investigated in more depth and independently of physicists
(see Ref. \cite{Vil91} and references quoted therein). From
point of view of the global differential geometry, at $n\in{\Bbb{Z}}$ the
monopole harmonics are
the {\it sections} (written in local form
on the bundle chart of coordinates $\vartheta,\varphi$ covering almost the
whole $S^2$) of the complex line
bundle with the Chern number $n$ over $S^2$, i. e., the ordinary ones are
the sections of trivial line bundle. If $n$ is not integer then, as far as is
known to the author, some geometric interpretation for monopole harmonics
is absent.

\end{document}